\documentclass{emulateapj}

\newcommand{\be}{  \begin{eqnarray} }
\newcommand{\ee}{  \end{eqnarray} }
\newcommand{\msun}{ M_{\odot}}

\shorttitle{Spectral Models of Black Hole Binaries}
\shortauthors{Davis and Hubeny}
\begin{document}
\title{A Grid of Relativistic, non-LTE Accretion Disk Models for Spectral Fitting of
	   Black Hole Binaries}

\author{Shane W. Davis\altaffilmark{1} and Ivan Hubeny\altaffilmark{2}}
\altaffiltext{1}{Department of Physics, University of California, Santa Barbara,
CA 93106}
\altaffiltext{2}{Steward Observatory and Department of Astronomy, University of Arizona, Tucson, AZ 85721}

\begin{abstract}

Self-consistent vertical structure models together with non-LTE radiative transfer should
produce spectra from accretion disks around black holes which differ from multitemperature
blackbodies at levels which may be observed. High resolution, high signal-to-noise
observations warrant spectral modeling which both accounts for relativistic effects, and
treats the physics of radiative transfer in detail.  In Davis et al. (2005) we presented
spectral models which accounted for non-LTE effects, Compton scattering, and the opacities due
to ions of abundant metals. Using a modification of this method, we have tabulated
spectra for black hole masses typical of Galactic binaries. We make them publicly available
for spectral fitting as an Xspec model.  These models represent the most complete realization
of standard accretion disk theory to date.  Thus, they are well suited for both testing the
theory's applicability to observed systems and for constraining properties of the black holes,
including their spins.

\end{abstract}


\section{Introduction}
 \label{intro}

It is commonly believed that some fraction of the X-ray emission from black hole X-ray
binaries (BHBs) is thermal emission from an optically thick accretion disk and that the
remaining non-thermal emission comes from a neighboring, optically thin corona. Spectral
states of BHBs are usually characterized by the relative strengths of these thermal and
non-thermal emission components (e.g. McClintock \& Remillard 2004). In the high/soft (or
thermal dominant) state, the thermal component dominates the spectral energy density (SED) in
the X-ray band, and a radiatively efficient disk is assumed to extend deep within the
gravitational potential of the black hole. In principle, spectral modeling of these sources
therefore provides both a means for testing accretion disk theory in the strong gravity near a
black hole and inferring the properties of the black hole itself.

There are two properties of black hole binaries which make them particularly well suited for
these goals.  First, precise, independent constraints on the binary properties such as the
accretor mass and the inclination of the binary are often available from light curve modeling
of the secondary star (see e.g. Orosz \& Bailyn 1997). These constraints reduce the number of
free parameters in the spectral models.  Such constraints are typically less common or more
uncertain for the super-massive black holes believed to power active galactic nuclei (AGNs).  
A second advantage of BHBs is that the same source may be observed accreting matter at
appreciably different rates.  This is simply not possible for most AGNs, in which the
timescales of interest are long compared with the human life span.  This property of BHBs has
already been used to draw interesting constraints on the nature of these accretion flows (e.g.
Kubota, Makishima, \& Ebisawa 2001; Gierli\'nski \& Done 2004, hereafter GD04).  These
investigations showed that for many BHBs in the high/soft state, the luminosity of the thermal
component $L$ scales as the fourth power of the color temperature $T_{\rm c}$ over a wide
range of accretion rates. This relation implies that the emitting area of the disk (and thus
the inner radius) changes by very little and that color temperature scales roughly linearly
with the effective temperature.

These results suggest that existing observations can already provide quantitative constraints,
and motivate the use and development of sophisticated spectral models. Almost all spectral
models which are commonly used to fit the thermal emission in BHBs are motivated in part by
the geometrically thin $\alpha$-disk (Shakura \& Sunyaev 1973; Novikov \& Thorne 1973). The
multicolor disk (MCD, Mitsuda et al. 1984) is one of the simplest and most widely used models,
and it generally provides an adequate fit to the soft thermal component in BHBs.  However, it
is not the best proxy for the standard $\alpha$-disk model as it ignores the no-torque inner
boundary assumption, neglects relativistic effects, and approximates the disk surface emission
as a blackbody.

There have been many efforts to generate spectral models which incorporate relativistic
effects or more of the thin disk physics (e.g. GRAD, Ebisawa, Mitsuda, \& Hanawa 1993;
DISKPN, Gierli\'nski et al. 1999; BMC, Borozdin et al. 1999, Shrader \& Titarchuk 1999). 
Currently, the most sophisticated fitting model of this
type is the KERRBB model (Li et al. 2005) which includes fully relativistic effects on the
disk structure (Novikov \& Thorne 1973) and photon transfer (Cunningham 1975).  One of the
main limitations of KERRBB (along with the MCD, GRAD, DISKPN, and similar models) is that it
approximates the potentially complex disk surface emission with a color corrected blackbody
\be \label{e:ccbb} I_{\nu}=f^{-4}B_{\nu}(fT_{\rm eff}) \ee where $I_{\nu}$ is the specific
intensity, $T_{\rm eff}$ is the effective temperature, $B_{\nu}$ is the Planck function, and
$f$ is the spectral hardening factor (color correction).

Several attempts have been made to calculate the vertical structure along with the resulting
emission in the local frame at the disk surface (Shimura \& Takahara 1995; Merloni et al.
2000; Davis et al. 2005, hereafter Paper I; Hui et al. 2005). Shimura \& Takahara (1995)
claimed that the color
corrected blackbody was a suitable approximation for most luminosities typical of high/soft
state BHBs.  Their results suggested that the spectral hardening factor was only weakly
dependent on luminosity and had a value of $\sim 1.7$.  In Paper I we calculated non-LTE
models with full radiative transfer including the bound-free opacity of metals and the effects
of Compton scattering.  Our conclusions were similar to those of Shimura \& Takahara (1995)
except that we were slightly less encouraged by the quality of the color corrected blackbody
approximation and found our results to be consistent with somewhat lower hardening factors. We
also showed that our full disk SEDs gave $L - T_{\rm c}$ relations which were qualitatively
consistent with those of sources discussed in GD04.

To date, none of these results have been directly implemented in spectral fitting models which
include relativistic effects. Shimura \& Takahara produced a table model capable of fitting
data, but this does not account for relativistic effects and is therefore of limited
applicability to BHBs. Our comparisons with GD04 in Paper I required us to generate artificial
spectra which we then fit with the MCD model.  A similar technique has been used to derive
hardening factors suitable for use in spectral fitting with KERRBB (Shafee et al., 2006).
However, it would also be useful to fit our models directly to the data and not need to rely
on the MCD model or KERRBB as intermediaries.  Motivated by these considerations, we have
implemented our accretion disk SEDs in a table model BHSPEC for the publicly available Xspec
spectral fitting package (Arnaud 1996).  In order to do this efficiently, we have made slight
modifications to the method we presented in Paper I and we discuss these changes below. In
section \ref{models}, we summarize our modified method for generating disk spectra and compare
the results to models which utilize the method presented in Paper I. In section \ref{discus}
we discuss the applicability of our spectral fitting models to data, their limitations, and
potential sources of error. Fits of these models to observations of BHBs will be presented
in a companion paper (Davis et al. 2006).

\section{Models}
 \label{models}

\subsection{Method}

Detailed discussions of our general method for calculating accretion disk spectra (hereafter
referred to as the direct method) and the related caveats are contained in Paper I and
references therein.  Here we only summarize the most salient features or those directly
related to our current modifications.  There are three main components to our calculation.  
First, we solve the fully relativistic one zone disk structure equations in the Kerr metric
(Novikov \& Thorne 1973; Page \& Thorne 1974; Riffert \& Herold 1985).  The parameters of this
one-zone calculation are the black hole mass $M$ and spin parameter $a_{\ast} \equiv a/M$, the
accretion rate, and the stress parameter $\alpha$. For most applications, $\alpha$ is the
constant of proportionality which relates the vertically averaged accretion stress to the
vertically averaged total pressure. For our Xspec model we parameterize the accretion rate by
$\ell \equiv L/L_{\rm Edd}$ where $L_{\rm Edd}=1.3 \times 10^{38} (M/M_{\odot})\,\,\rm erg\, s^{-1}$ is 
the Eddington luminosity for completely
ionized H.  To convert from accretion rate to $\ell$ we assume an efficiency $\eta$ which is the
ratio of the total radiative luminosity emitted from the disk surface to the rate of rest mass
energy supplied by mass accretion.  For disks with a no-torque inner boundary condition,
$\eta$ depends only on $a_{\ast}$. For torqued disk models we parameterize the magnitude of
the torque by the change in $\eta$ implied by the increased radiative luminosity due to work
done by the torque.

With the direct method, the next step is to divide the disk into a set of concentric annuli,
and to compute their local vertical structure and radiative transfer using the program TLUSTY
(Hubeny \& Lanz 1995). The code was originally designed for model stellar atmospheres, and had
a distinct variant TLUSDISK for computing vertical structure of accretion disks; the latest
versions (e.g. version 200\footnotemark) represent a universal code for both stellar
atmospheres and accretion disks.

\footnotetext{http://tlusty.gsfc.nasa.gov}

The only additional parameter specifying a given annulus is its radial coordinate $r$. The
one-zone model from the previous step allows us to compute for a given $r$ the actual input
parameters for the annulus, which are the effective temperature $T_{\rm eff}$, the column mass
at the midplane $m_0$ ($m_0=\Sigma/2$ where $\Sigma$ is the total disk surface density), and 
gravity parameter $Q=g/z$ which is the constant of
proportionality between the local gravity $g$ and height above the midplane $z$ 
(Hubeny \& Hubeny, 1998).
Finally, we calculate the integrated disk spectrum seen by an observer at infinity by
computing photon geodesics in a fully general-relativistic spacetime (Agol 1997).

The main drawback of this direct method is that it requires a minimum of 20-30 disk annulus
calculations to accurately resolve the variation with radius of the surface emission over the
energy band of interest (0.1 keV and above). This creates difficulties in the production of a
fittable model. Due to the complexity of the system of equations being solved, the convergence
rate of our disk annulus calculations is not 100\%.  Therefore, the calculations cannot be
entirely automated, making direct computation of the model on each fitting step infeasible.  
Any realistic scheme of spectral fitting with these models requires some method of
interpolation on a precomputed set of data. So, we have chosen to implement our fitting model
as an Xspec table model (Arnaud 1996).  This frequency-by-frequency interpolation method
requires a reasonably fine grid resolution in the thin disk model parameters to accurately
represent the interpolated spectrum. We would like to explore several parameters over a wide
range of values and we find that these considerations require us to compute hundreds of
thousands of disk spectra. The method described above would then require millions of disk
annulus calculations. However, the need for human intervention in the case of unconverged
annuli sets a practical upper limit on the number of disk annulus calculations that can be
performed even if unlimited computing resources were available.

To satisfy these constraints, we make a simple modification to our scheme which greatly
reduces the total number of calculations which are needed.  The modification relies on the
fact that each annulus is determined by only three parameters in the method outlined above.
Calculation of a large grid of disk spectra would therefore result in the computation of many
annuli with very similar parameters.  So, instead of computing the spectra of each of these
annuli directly, we have computed a table of $\sim 800$ annuli in the parameters $\Sigma$,
$T_{\rm eff}$, and $Q$ from which we can interpolate spectra for a specific set of values.
Expansion of the grid is ongoing, but currently the table of model annuli (which is not a
square grid) covers a range of $\log T_{\rm eff}$ from 5.0 to 7.4 in steps of 0.1; $\log Q$
from -4.0 to 9.0 in steps of 1.0; and $\log m_0$ at 2.5, 2.75, and from 3.0 to 6.0 in steps of
1.0.

\subsection{Table of Annuli Spectra}
\label{toa}

The physics of the disk vertical structure and radiative transfer has been discussed in detail
elsewhere (see Paper I and references therein) so we only summarize our results here. We make
the standard assumptions that the annuli are time steady and their structure depends only on height.  
Following Hubeny et al. (2001), our annuli models include Compton scattering, bound-free and
free-free opacity of H, He, and most important metals -- C, N, O, Ne, Mg, Si, S, Ar, Ca, and
Fe.  In non-LTE calculations, hydrogen is represented by a 9-level atom; the hydrogenic ions
as 4-level atoms, and all ions of other metals by one-level atoms. Bound-free transitions
include the standard outer shell ionization processes, an approximate treatment of Auger
(inner shell)  processes, as well as collisional ionization processes. Free-bound processes
include the radiative, dielectronic, and three-body recombination processes. The opacities and
radiative transition rates due to bound-bound processes (spectral lines) are neglected. The
models additionally assume that dissipation is locally proportional to density and all
vertical energy flux is transported radiatively. We neglect the contribution of the magnetic
field forces to the hydrostatic equilibrium, and we do not include irradiation at the upper
boundary. These assumptions, along with the elemental abundances, $\Sigma$, $Q$, and $T_{\rm
eff}$ completely determine the vertical structure and emitted spectrum.

In the effectively optically thick limit, the spectrum depends most strongly on $T_{\rm eff}$. The
gravity parameter $Q$ is also important in determining both the strength of edges and the
relative importance of scattering and absorption opacity.  The spectra are only weakly
dependent on $\Sigma$ as long as the annuli remain effectively optically thick.  This weak dependence on
$\Sigma$ in the effectively optically thick limit is analogous to standard results from the spectral
modeling of stellar atmospheres. In stars, essentially all of the energy generation occurs deep
within the stellar interior so that the radiative flux through the envelope is constant.  
Thus, the spectra are independent of the total column mass used in the calculation, provided
that the lower boundary is chosen to be at sufficiently high optical depth that the photons
come into LTE.

\begin{figure}
\includegraphics[width=0.5\textwidth]{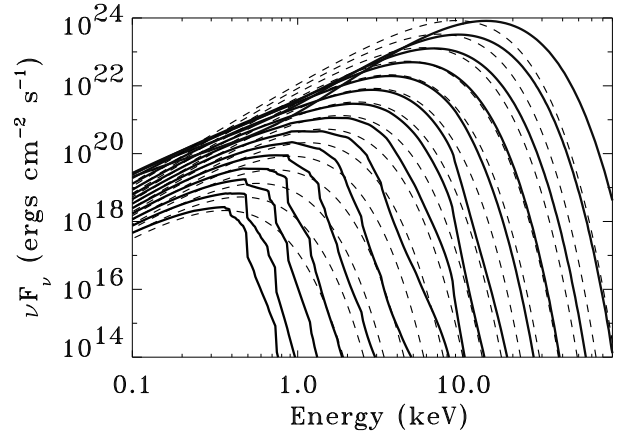}
%\plotone{f1.eps}
\caption{
Specific intensity of 15 annuli viewed at an inclination of $55^{\circ}$ relative to the
surface normal vector.  The annuli are evenly spaced in $\log T_{\rm eff}$ between
$\log T_{\rm eff}=5.8$ and 7.2 where $T_{\rm eff}$ has units of Kelvin.  This change of 0.1 dex 
between spectra corresponds to the resolution used in our table of annuli.  All annuli
are computed for $\Sigma=2 \times 10^4$g/cm$^2$ and $Q=10^6$s$^{-2}$.  For comparison, we
also plot the color corrected blackbody spectra (eq. \ref{e:ccbb}, dashed curves) with 
fixed $f=1.7$ at the same values of $T_{\rm eff}$.
\label{f:flux}}
\end{figure}

In Figure \ref{f:flux} we plot the spectra from annuli with the same values of $\Sigma$ and
$Q$ but in which $T_{\rm eff}$ is varied.  At moderate to high $T_{\rm eff}$ the spectral
formation is dominated by modified blackbody effects and Comptonization. The resulting
spectral shape may be approximated to some degree by a color-corrected blackbody, but requires a spectral
hardening factor which increases with increasing $T_{\rm eff}$. A single value of $f=1.7$ is not
sufficient for all annuli. At lower $T_{\rm eff}$ the
effects of the absorptions edges are significant and the color corrected blackbody provides 
a poor approximation.

\begin{figure}
\includegraphics[width=0.5\textwidth]{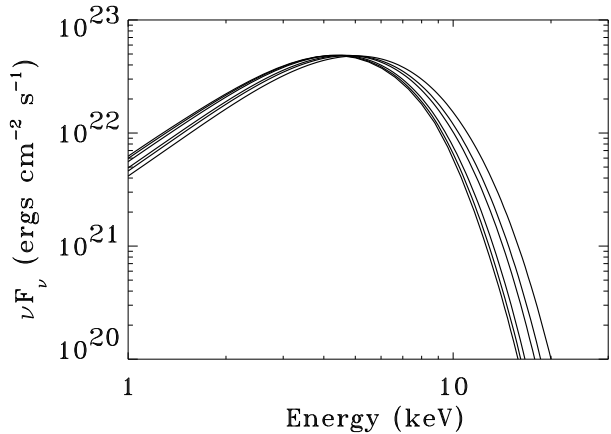}
%\plotone{f2.eps}
\caption{
Specific intensity of 6 annuli viewed at an inclination of $55^{\circ}$ relative to the
surface normal vector.  From right-to-left, the annuli correspond to $\Sigma=632, 1.12 
\times 10^3, 2 \times 10^3, 2 \times 10^4, 2 \times 10^5$, and $2 \times 10^6$g/cm$^2$.  
These changes in $\Sigma$ correspond to the resolution
in our table of annuli.  All annuli are computed for $T_{\rm eff}=7.9 \times 10^6$K and 
$Q=10^7$s$^{-2}$. The three models with the
largest $\Sigma$ are all effectively thick so that spectrum changes only slightly over two
orders of magnitude in $\Sigma$.  When $\Sigma$ drops below $\Sigma \sim 10^4$g/cm$^2$, 
spectra begin to harden more rapidly as they become increasingly effectively thin.
\label{f:m0}}
\end{figure}

As the annuli start to become marginally effectively optically thin, the spectra become increasingly
sensitive to $\Sigma$ as seen in Figure \ref{f:m0}. In these cases, the region of spectral
formation extends deep into the atmosphere where densities depend more sensitively on
$\Sigma$.  The scale height is only weakly dependent on $\Sigma$, so the density drops as
$\Sigma$ decreases. Therefore, the thermalization surface moves even deeper into the
atmosphere where the temperatures are higher resulting in a harder spectrum. As the annuli
become increasingly effectively optically thin the temperature distribution becomes increasingly
isothermal and Compton scattering completely determines the radiative equilibrium.  In the
effectively optically thin limit, the number of seed photons produced by the disk scales
roughly proportional to $\int \rho^2 d z \simeq \Sigma^2/h$.  A decrease in $\Sigma$ results
in a decrease in the number of seed photons which must be compensated by an increase in
average photon energy to provide the same surface flux. Thus, in the effectively optically thin limit,
the spectrum rapidly hardens as $\Sigma$ decreases.  The total number of photons emitted also
depends on $Q$ and $T_{\rm eff}$ via the scale height $h$. At fixed $\Sigma$ the number of
seed photons decreases for increasing $T_{\rm eff}$ or decreasing $Q$ since $h \propto T_{\rm
eff}^4/Q$.  Therefore, the spectral hardening depends sensitively on all three parameters in the
effectively optically thin limit.

\subsection{Accretion Disk SEDs}
\label{ads}

In order to generate a full accretion disk SED, we require a model for the annuli parameters
$\Sigma$, $Q$, and $T_{\rm eff}$ as a function of radius. Our spectrum-generating software
first sets up discretized radial coordinates $r_i$. The one-zone model then computes basic
parameters $\Sigma(r_i)$, $T_{\rm eff}(r_i)$ and $Q(r_i)$ for each radius $r_i$. The local
emergent specific intensity for each annulus is then obtained by interpolating from the table
of annuli spectra. Finally, the integrated disk SED for a given inclination is computed with a
general-relativistic photon transfer function using KERRTRANS (Agol 1997). Provided our
assumptions about vertical structure apply, the annuli discussed in section \ref{toa} can be
used with {\it any} one-zone disk model which self-consistently determines these parameters.
In this paper, we confine our attention to standard thin disk models in which the vertically
averaged stress $\tau_{r \phi}$ is given by \be \tau_{r \phi}=\alpha P \label{e:stress} \ee
where $P$ is the vertically averaged {\it total} (gas plus radiation) pressure and $\alpha$ is
a constant. This choice of stress prescription determines $\Sigma$ for a given accretion rate
and radius.

The interpolation scheme is a potential source of error in our method. Due to atomic features,
we must interpolate frequency-by-frequency. However, if we interpolate using the specific
intensity $I_{\nu}$, the interpolated spectrum is typically a poor approximation at
frequencies just above the spectral peak due to the exponential dependence on frequency in the
high energy tails. In order to maximize the accuracy of our interpolation scheme for a given
resolution in $T_{\rm eff}$, we account for this exponential dependence by interpolating in
terms of a color-corrected brightness temperature \be T_{\rm B} \equiv \frac{h \nu}{f k_{\rm
B}}\left[\log\left( 1 + \frac{2 h \nu^3}{c^2 f^4 I_{\nu}} \right) \right]^{-1} \ee instead of
the specific intensity.  Here, $\nu$ is the photon frequency, $k_{\rm B}$ is Boltzman's
constant, $h$ is Planck's constant, $c$ is the speed of light, and $f$ is a spectral hardening
factor which accounts for the fact that average photon energy is higher than in the blackbody
case. Through trial and error, we found $f=2$ worked well for a wide range of annuli spectra,
but the spectra do not depend sensitively on this choice. This brightness temperature provides 
a one-to-one mapping with the specific intensity.

\begin{figure}
\includegraphics[width=0.5\textwidth]{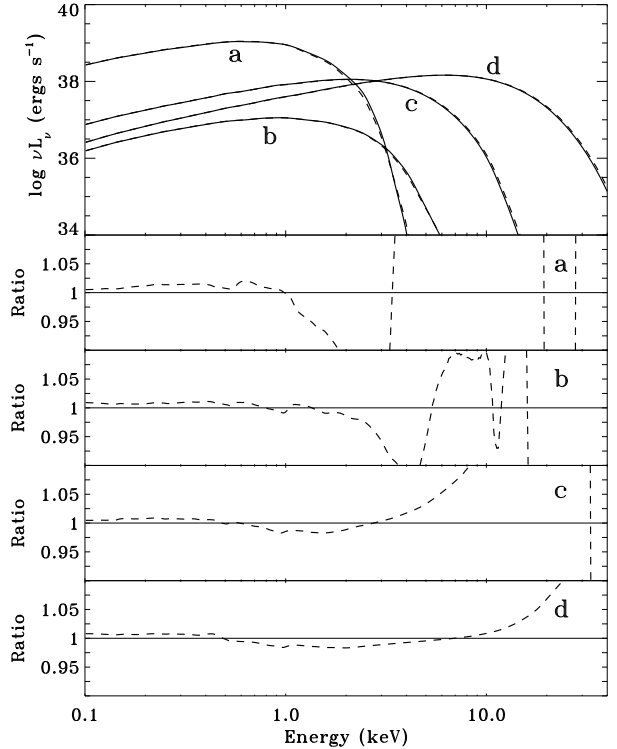}
%\epsscale{.8}\plotone{f3.eps}\epsscale{1}
\caption{
Specific intensity viewed at an inclination of $55^{\circ}$ for four disk models: 
model a with $M=300 \msun$, $a_{\ast}=0.0$, $\alpha=0.01$, and $\ell=0.1$; 
model b with $M=10 \msun$, $a_{\ast}=0.0$, $\alpha=0.1$, and $\ell=0.03$; 
model c with $M=10 \msun$, $a_{\ast}=0.0$, $\alpha=0.1$, and $\ell=0.3$; and 
model d with $M=10 \msun$, $a_{\ast}=0.998$, $\alpha=0.1$, and $\ell=0.3$.  
In the upper panel the solid curves represent the models
in which the annuli spectra are calculated directly and the dashed curves represent
the models in which the annuli spectra are interpolated from the table.  In the lower panel,
we show the ratio of the spectra from the interpolated method to the spectra from the
exact method.
\label{f:comp}}
\end{figure}

The effectiveness of this method decreases when the shape of the spectrum deviates
significantly from Planckian. Therefore, the interpolation method is generally somewhat less
accurate for those annuli with $T_{\rm eff} \lesssim 10^6 \, \rm K$ which have prominent
absorption edges in their spectrum. In this range the interpolation tends to underpredict the
flux in the spectral tail.  A second difficulty occurs at high temperatures when the annuli
start to become effectively thin.  As discussed in section \ref{toa}, the spectra of annuli
harden rapidly with increasing $T_{\rm eff}$.  Our grid has difficulties resolving this
rapid, non-linear hardening of the spectra and the interpolation tends to produce a
spectrum which is too hard. As a result, the high energy tails of the interpolated full disk
spectra are slightly harder than those in which each annulus is calculated directly.

Both effects can be seen in Figure \ref{f:comp}.  We show SEDs from four representative
models:  model a with $M=300 \msun$, $a_{\ast}=0.0$, $\alpha=0.01$, and $\ell=0.1$;  model b with $M=10
\msun$, $a_{\ast}=0.0$, $\alpha=0.1$, and $\ell=0.03$; model c with $M=10 \msun$, $a_{\ast}=0.0$, $\alpha=0.1$,
and $\ell=0.3$; and model d with $M=10 \msun$, $a_{\ast}=0.998$, $\alpha=0.1$, and $\ell=0.3$.  The solid
curves represent models calculated with the direct method and the dashed curves show the
models in which the annuli spectra are interpolated using the table. In models a and b the
interpolated spectra tend to underestimate the flux just above the peak due the non-Planckian
shape of the spectra.  For models c and d, the hottest annuli are becoming effectively
optically thin and spectral hardening is not as well resolved by our table. As a result, the
interpolation method overpredicts the flux just above the spectral peak. At or below the peak
frequency, the two methods agree to within 1-2\%.  Just above the spectral peak the
disagreement begins to increase to 5-10 \% as the spectrum decreases to a decade below the
peak. The worst disagreement occurs far out in the high energy tail where the photon
statistics will be low, but the discrepancies at energies slightly above the peak may still 
influence fits to high signal-to-noise data. The shape of the tail is typically determined only 
by the contribution of the hottest annulus, and these discrepancies represent the limit of
the accuracy which can be achieved with interpolation method for the current resolution of
our table of annuli. Therefore, improving the resolution in log $T_{\rm eff}$ is a goal
for future efforts. The implications of these discrepancies for spectral fitting 
are discussed further in section \ref{discus}.

\begin{figure}
\plotone{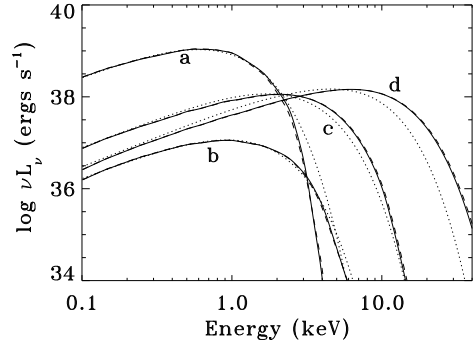}
\caption{
Specific intensity viewed at an inclination of $55^{\circ}$ for four representative disk 
models. The model labels are defined in the caption of Figure \ref{f:comp} and in the text. 
The solid curves represent the models in which the annuli spectra are calculated directly and 
the dashed curves represent the models in which the annuli spectra are interpolated 
from the table. The dotted curves are the best fit color corrected blackbody spectra.  The fitting
procedure is described in section \ref{ads}.
\label{f:comp2}}
\end{figure}

In Figure \ref{f:comp2} we compare the direct method SEDs (solid curves) with
fully-relativistic models which represent the disk surface emission using equation
(\ref{e:ccbb}) (dotted curves). We fit each of the full disk SEDs with these color corrected
blackbody spectra and found $f=1.65$, 1.46, 1.65, and 1.81 for models a, b, c, and d
respectively.  The photon counts spectra were fit to photon energies above 0.1 keV assuming a
constant effective area.  It is clear that the interpolated spectra (dashed curves) provide
much better approximations to the direct method calculation than those using equation
(\ref{e:ccbb}). These color-corrected blackbody models provide a particularly poor approximation
for models c and d.  For these models, the hottest annuli spectra require values of $f$
considerably larger than those of the cooler, effectively thicker annuli at larger radius.
Spectra calculated using a single value for color correction at all radii cannot adequately
model the whole SED above 0.1 keV.

\section{Spectral Fitting: Applications and Limitations}
\label{discus}

We have used the interpolation method described in section \ref{models} to tabulate a large
number of artificial disk SEDs. These have yielded Xspec table models
(collectively referred to as BHSPEC) which are publicly available\footnotemark. The different
table models cover different ranges and combinations of the parameters of interest. The
primary model parameters are $M$, $\ell$, $a_{\ast}$, $\alpha$, and the inclination $i$, but we
have also produced models which allow the metallicity or the magnitude of the torque on the
inner edge of the disk to vary (Agol \& Krolik 2000). The Xspec model also has a normalization 
$N$, defined so that
$N=(10 {\rm kpc}/D)^2$ where $D$ is the distance to the binary. The dependence of disk spectra 
on these parameters is discussed in detail in Paper I and further
information can be found in Davis et al. (2006) and the documentation accompanying the models.

In principle, fits of BHSPEC to high signal-to-noise data can allow for very precise estimates
of the fitting parameters.  However, there are several concerns which limit the accuracy of
the model and the types of data to which it can be reliably applied.
From Figure \ref{f:comp}, we conclude that the interpolation method produces a potentially
significant error at photon frequencies just above the spectral peak. In order to understand
how this error might affect spectral modeling, we performed fits to artificial data
generated with model c from Figure \ref{f:comp}.  Specifically, we used an {\it RXTE}
Proportional Counter Array (PCA) response matrix and assumed a 2000 second exposure to
generate artificial datasets for the model computed with the exact method.  We then fit
these data over the 3-20 keV band with BHSPEC. Since the shape of the BHSPEC spectrum is
degenerate to different combinations of the fitting parameters (see Davis et al. 2006 for
further discussion), sources with precise and reliable estimates for some parameters will
provide better constraints for the unknown parameters. In order to simulate applications to
such sources, we assumed that $D$ and $M$ are known to 10\% and only allowed $M$ and the
normalization to vary over the corresponding range in our fit.

\begin{deluxetable*}{lccccc}
\tablecolumns{6}
\tablecaption{Interpolation Method Error Estimate\label{tbl1}}
\tablewidth{0pt}
\tablehead{
 &
\colhead{$M$\tablenotemark{a}} &
\colhead{$i$} &
\colhead{$\ell$} &
\colhead{$D$\tablenotemark{a}} &
\colhead{$a_\ast$} \\
 &
\colhead{$(M_{\odot})$} &
\colhead{$(^\circ)$} &
&
\colhead{(kpc)} &
}

\startdata
Model &
10 &
70 &
0.3 &
5 &
0 \\
BHSPEC Fit\tablenotemark{b} &
$9.49^{+0.14}_{-0.08}$ &
$67.8^{+0.4}_{-3.4}$ &
$0.2364^{+0.0044}_{-0.0020}$ &
$4.5^{+0.37}_{-0}$ &
$0.200^{+0.005}_{-0.2}$\\

\enddata
\tablenotetext{a}{To simulate constraints for realistic spectral modeling, we  assume $D$ and $M$ are `known' 
to 10\% and limit the allowed range of the normalization and $M$ accordingly.}
\tablenotetext{b}{$\chi^2_{\nu}=56.4/40$.}
\tablecomments{All uncertainties are 90\% confidence for one parameter.}

\end{deluxetable*}

The fit results are reported in Table \ref{tbl1}.  Even for 2000 sec, the signal-to-noise is
quite high, and we find a relatively poor fit with $\chi^2_\nu=56/40$.  The assumed
constraints are important since the best-fit $D$=4.5 kpc, the limit we have imposed.  (In
this case, the best-fit $M$ lies well within the assumed 10\% uncertainty range.) If no
limits are imposed, the quality of fit will be better for lower $D$, but the best-fit
parameters can differ significantly from those assumed in generating the model.  Therefore,
we caution against fitting data from sources where independent constraints are not
available, particularly with {\it RXTE} PCA data.  Except for $a_\ast$, none of the exact
model parameters lie in the 90\% confidence ranges of the best-fit parameters.  These
confidence ranges are therefore not to be trusted. Nevertheless, we find that the best-fit
values of i and M are correct to within 5\%, and $\ell$ is correct to about 20\%. A
difference $\Delta a_\ast \sim 0.2$ at low $a_\ast$ corresponds to only a modest change in
the disk inner radius. Hence while these errors should be kept in mind when interpreting fit
results (especially with high signal-to-noise PCA data), we do not consider them
particularly discouraging.  We have chosen {\it RXTE} for this comparison partly because it
has one of the hardest energy bands among the X-ray detectors for which the BHSPEC model
would be useful, and it therefore provides approximate `upper limits' on the errors we
expect due to our interpolation scheme.  Better agreement can be expected from X-ray
observatories with sensitivity at softer photon energies for which our interpolation method
provides a better approximation.

There are also important practical limitations for this method.  Perhaps the most
important are the difficulties incorporating irradiation of the disk surface by a corona, a
central star in the case of neutron stars or white dwarfs, or self-irradiation by the disk
itself.  In these cases, TLUSTY can also be used to calculate annuli with surface irradiation.
So we can construct full disk SEDs using the direct calculation method outlined in Paper I,
but it may be difficult to approximate the surface emission in a way that is easy to
parameterize in a table.  Even a simple prescription which assumes either a diluted blackbody
or powerlaw form for the irradiating spectrum would introduce at least two more parameters to
our table and greatly increase the number of annuli which would need to be computed.  Nor is
it clear that such simple spectral shapes would be adequate.  For example, if one assumes the
non-thermal emission in a BHB is due to upscattered emission from the disk surface, a simple
power law form would overestimate the irradiating flux at lower photon energies.  In the case
of self-irradiation where the irradiating photons are coming from other regions of the
accretion disk surface, the spectrum will not generally be well approximated by either of
these simple spectral shapes.

The neglect of irradiation limits the types of
observations for which BHSPEC can be expected to be applied self-consistently.  It is best
suited for high/soft state SEDs with a low fraction of the bolometric flux in the
non-thermal component. It is difficult to quantify how small this fraction needs to be until
further work examining irradiated disks is completed.  However, a reasonably large sample of
{\it RXTE} observations of BHBs exists with less than 10-15\% of the bolometric flux
inferred to be in the non-thermal component (GD04). In these sources, we expect that coronal
irradiation will have limited impact and our spectral models may still provide a good
approximation. BHSPEC has been applied to spectral fitting of a subset of BHBs in this
sample and these fits will be presented in a companion paper (Davis et al. 2006).

The geometrically thin $\alpha$-disk suffers from a number of other potential difficulties,
and may fail to adequately approximate the structure and dynamics of real magnetohydrodynamical
accretion flows.  Nevertheless, it is among the simplest models to compute, and we have 
implemented it with the 
expectation that it may adequately approximate the spectra of more complex accretion flows.
The procedures outlined in section \ref{models} are flexible and can be adapted to consider
more general models, if necessary. Now that a table of annuli has been constructed, it is
straightforward to use
it with other one-zone models for the disk structure.  For example, one can substitute
$P_{\rm gas}$ for $P_{\rm tot}$ in equation (\ref{e:stress}) and construct so-called
$\beta$-disk models.  Such models have the advantage that they are not subject to the 
`viscous' and thermal instabilities of the radiation pressure dominate regime (Piran 1978)
which plague the $\alpha$-disk models upon which BHSPEC is based.

Another potential problem is that the scale height $H$ of the
disk becomes comparable to $R$ for sufficiently high $\ell$. For models with $\ell=0.3$,
the height of the photosphere in our models is less than $0.1 R$ at all radii, and 
we expect the thin disk
model remains a reasonable approximation.  At still higher $\ell$ the effects of advection
and radial pressure gradients can no longer be ignored and slim disk models (e.g. Abramowicz
et al. 1988) should be considered.  In principle, one could also use this method to calculate
$\Sigma$, $Q$, and $T_{\rm eff}$ as a function of radius in such models.  Preliminary work
suggests that slim disk models are reasonably well approximated by their thin disk counterparts
for accretion rate up to the Eddington rate as long as a $\beta$-disk prescription is
utilized. However, such calculations may still need to address the effects of a three
dimensional trapping surface on the vertical structure.  The relativistic photon transfer
might also require modifications to account for the fact that photons leave the disk surface
at high latitudes relative to the disk midplane.  Therefore, fit results should be interpreted
cautiously in the $\ell \sim 1$ regime.

\footnotetext{http://heasarc.gsfc.nasa.gov/docs/xanadu/xspec/models/bhspec.html or
http://www.physics.ucsb.edu/$\sim$swd/xspec.html}

\acknowledgements{We thank Omer Blaes, Chris Done, and the referee, David Ballantyne, for helpful 
suggestions and careful reading of the manuscript. We also thank Julian Krolik for useful 
discussions and Eric Agol for making his relativistic
transfer code (KERRTRANS) publicly available.  This work took advantage of software and
documentation for Xspec which was provided by Keith Arnaud. Part of this work was carried out while S.D. 
was hosted by  the Kavli Institute for Theoretical Astrophysics. This work was supported in part by NASA 
grant NAG5-13228 and NSF grant PHY99-0794.}

\end{document}